\documentstyle[12pt]{article}
\def\beq{\begin{equation}}
\def\eeq{\end{equation}}
\def\beqa{\begin{eqnarray}}
\def\eeqa{\end{eqnarray}}
\def\MeV{\nobreak\,\mbox{MeV}}
\def\GeV{\nobreak\,\mbox{GeV}}
\def\fm{\nobreak\,\mbox{fm}}
\def\qbar{\overline{q}}
\def\ubar{\overline{u}}
\def\dbar{\overline{d}}
\def\sbar{\overline{s}}
\def\bra#1{\langle #1|}
\def\ket#1{| #1\rangle}
\def\nucbra{\bra{ N}}
\def\nucket{\ket{ N}}

\begin{document}

%%%%%%%%% %%%%%%%%% %%%%%%%%% %%%%%%%%% %%%%%%%%% %%%%%%%%% %%

\thispagestyle{empty}
\setcounter{page}{0}

\noindent {\it to appear in Phys.\ Lett.\ B} \hfill U. of MD
PP \#93--217

\hfill DOE/ER/40762--011

\vspace*{48pt}

\centerline{\Large\bf Just How Strange? Loops, Poles and }
\centerline{\Large\bf the Strangeness Radius of the Nucleon}
\vspace{30.4pt}
\centerline{\sc Thomas D. Cohen, Hilmar Forkel and Marina
Nielsen\footnote{Permanent address: Instituto de F\'\i sica,
Universidade de S\~ao Paulo, 01498 - SP- Brazil.}}
\vspace{7.2pt}
\centerline{\it Department of Physics, University of Maryland}
\centerline{\it College Park, MD 20742}
\vspace{14.4pt}
\centerline{August, 1993}
\vspace{21.6pt}

\begin{quote}\hspace*{0.5cm} We consider a simple model for the
strangeness radius of the nucleon. The model is based on vector
meson dominance (VMD) and $\omega - \phi$ mixing in addition to
a kaon cloud contribution. We find that the VMD contribution is
similar in magnitude and of the same sign as the kaon contribution
to the Sachs strangeness radius and is significantly larger than
the kaon contribution to the Dirac radius.
\end{quote}
\vspace*{120pt}
\eject

In a recent calculation  by Musolf and Burkardt \cite{mu}, the
nucleon's mean square strangeness radius was evaluated in a
simple picture: the entire
contribution to the matrix elements $\nucbra\sbar\gamma_\mu
s\nucket$ was attributed to the kaon cloud of the nucleon and
estimated in a one-loop calculation.   This calculation was
intended to complement
Jaffe's model \cite{ja}, where the strangeness radius (and
magnetic moment)
is obtained by using a 3-pole fit for the spectral function of the
isoscalar form factors of the nucleon, in which one of the poles is
assumed to be the physical $\phi$ meson. We will refer to Jaffe's
approach as the ``pole picture''.

The results of these two pictures differ significantly. First of
all, loop and pole predictions for the strangeness radius have
opposite sign (the sign of the loop calculation agrees with the
Skyrme model estimates \cite{ps}). Furthermore, the magnitude of
the loop prediction for the strangeness radius is about 5 times
smaller than the pole result for the Sachs \cite{sachs} form and
about 20 times smaller for the Dirac radius.

In order to set scales we note that the Sachs charge radius of the
neutron is $<r^2_n>_{Sachs}\simeq -0.12\fm^2$ \cite{hohler} and
that the pole calculation gives a Sachs strangeness radius  of
the same order of magnitude :
$<r^2_s>_{Sachs}\simeq (0.14\pm 0.07)\fm^2$. The Dirac charge
radius of the neutron, on the other hand, is  $<r^2_n>_{Dirac}
\simeq 0.01\fm^2$, to be compared with $<r^2_s>_{Dirac}\simeq
(0.16\pm 0.06)\fm^2$ obtained in the pole calculation.

What conclusions can be inferred from this disagreement between
the two calculations? If the pole result is correct, then the
nucleon has a surprisingly large strangeness radius, meaning
that the contribution of strange quarks to the structure of the
nucleon is rather important. The nucleon is the lightest state
with nonzero baryon number. Why should it choose a quark
configuration for which $\nucbra\sbar\gamma_\mu s\nucket$ is so
large? On the other hand, if the loop result is correct, one
might argue that the strange quark contribution to the nucleon
structure is generally small, as one would naively expect from
the OZI rule. However, there are experimental indication for
non-negligible contributions from the strange quark sector to
other properties of the nucleon as, for instance, the scalar
strange quark density \cite{gasser}, the strange quark axial-vector
form factor seen in the elastic $\nu p/\nu p$ cross section
\cite{ahrens}, and the strange quark contribution to the proton
spin, measured by the EMC \cite{ashman}.

In the light of these questions, it seems quite probable to us
that the truth lies somewhere between the results of the two
calculations. It is clear that the physical pictures underlying
the models in refs.~\cite{mu} and \cite{ja} are quite different.
It is not immediately clear how to combine the ``pole-ish''
physics of ref.~\cite{ja} with the ``loopy'' physics at ref.~
\cite{mu}. In particular, it seems likely that one can not simply
add the two results on the grounds that this will lead to some
double counting.
It is the purpose of the present note to establish a link between
the pole and loop pictures, by combining the vector meson dominance
(VMD) model \cite{sa} in the $\omega$ and $\phi$ sector $(Y=T=0$,
$J^{PC}=1^{--})$ with the loop calculation. The model we propose
dynamically incorporates both the physics of the vector mesons
and kaons, and thus avoids double counting. In contrast to the
approach of ref.~\cite{ja}, it models the $\phi$-nucleon coupling
explicitly, in the VMD framework.

Were the $\omega$ and $\phi$ vector mesons pure $|\sbar \gamma_
\mu s>$ and $(|\ubar\gamma_\mu u> + |\dbar\gamma_\mu d>)/
\sqrt{2}$ states respectively, then the strange current,
$J_\mu^s=\sbar \gamma_\mu s$ could interact with the nucleon
only through the $\phi$ meson. In this case, in the VMD picture,
we could write for the strangeness form factor
\beq
F^s(q^2)=\frac{m_\phi^2}{m_\phi^2-q^2}F_i^s(q^2)\; ,
\label{1}
\eeq
where $F_i^s(q^2)$ is the intrinsic strangeness form factor of
the nucleon which one might associate with the kaon loop
calculation. Note that eq.~(\ref{1}) is independent of both
the $\phi-\gamma$ and $\phi$-nucleon couplings. As usual in
VMD models, they cancel as a consequence of  charge
normalization \cite{sa}. From eq.~(\ref{1}) it is clear that
the strangeness radius would be given by
\beq
<r^2_s> = \left.6\frac{\partial F^s(q^2)}{\partial q^2}
\right|_{q^2=0} = \left.6\left( \frac{1}{m_\phi^2} F_i^s(0)
+ \frac{\partial F^s_i(q^2)}{\partial q^2}\right|_{q^2=0}\right)\; .
\label{2}
\eeq
Since $F^s_i(0)=0$, the result is exactly the same as in the
loop calculation of ref.~\cite{mu}. However, in nature $\omega$
and $\phi$ are not pure $(|\ubar\gamma_\mu u> + |\dbar\gamma_
\mu d>)/\sqrt{2}$ and
$|\sbar \gamma_\mu s>$ --- they mix. The physical $\omega$ and
$\phi$ vector mesons are linear combinations of pure $|\sbar
\gamma_\mu s>$ $\; (\phi_0)$ and $(|\ubar\gamma_\mu u> + |
\dbar\gamma_\mu d>)/\sqrt{2}$ $\; (\omega_0)$ states, as in
ref.~\cite{jain} :
\beqa
|\omega> = \cos\epsilon\,|\omega_0> - \sin\epsilon\,|\phi_0>\ ,
\nonumber\\*[7.2pt]
|\phi> = \sin\epsilon\,|\omega_0> + \cos\epsilon\,|\phi_0>\ ,
\label{states}
\eeqa
therefore, the $\omega$ meson can contribute to the strangeness
radius.

The mixing angle, $\epsilon$, is known to be small and we will
use the value of ref.~\cite{jain} : $\epsilon=0.053$, which is
deduced from the decay of the $\phi$ meson into $\pi + \gamma $
and is consistent with its decay  into $\pi^+ +\pi^- +\pi^0$.
Our results are clearly sensitive to the value of the mixing angle.

Let us now generalize the VMD ideas to take the $\phi - \omega$
mixing into account. Consider the currents
\beqa
J_\mu^o &=& \frac{1}{2}(\ubar\gamma_\mu u + \dbar\gamma_\mu d)\ ,
\nonumber\\*[7.2pt]
J_\mu^s &=& \sbar\gamma_\mu s\ .
\label{cur}
\eeqa
In the VMD picture, one can express the form factors as the
product of a vector meson propagator (appropriately normalized)
and an intrinsic form factor, describing the vector meson coupling
to the nucleon:
\beq
\left(\begin{array}{c}
F_n^o(q^2)\\
F_n^s(q^2)
\end{array} \right) = \hat{f}_V(q^2,\epsilon) \left(\begin{array}{c}
F_{ni}^o(q^2)\\
F_{ni}^s(q^2)
\end{array}\right)\ ,
\label{form}
\eeq
where $n$ stands for Dirac ($n=1$) and Pauli ($n=2$) form factors
and $\hat{f}_V(q^2,\epsilon)$ is a matrix that gives the
contribution of both $\omega$ and $\phi$ mesons to the
considered currents. For a vanishing mixing angle, $\epsilon$,
we have
\beq
\hat{f}_V(q^2,0) = \left(\begin{array}{cc}
\frac{m^2_\omega}{m_\omega^2-q^2} & 0\\
0 & \frac{m_\phi^2}{m_\phi^2-q^2}
\end{array} \right)\ ,
\label{form0}
\eeq
and eq.~(\ref{1}) is recovered.

For $\epsilon\neq 0$ the above matrix eq.~(\ref{form0}) will
be rotated from its diagonal form,
\beq
\hat{f}_V(q^2,\epsilon) =
\hat{C}(\epsilon)\hat{f}_V(q^2,0)\hat{C}^{-1}(\epsilon)\; ,
\label{rot}
\eeq
where the matrix $\hat{C}$ gives the coupling of the mesons in
eq.~(\ref{states}) with the currents in eq.~(\ref{cur}).

Following Jaffe's prescription for the vector-meson current
couplings \cite{ja}, each quark $q_k$ couples to the current
$\qbar_k\gamma_\mu q_k$ with the same strength $K$ and does
not couple at all to currents of different flavor. As a result
we can write
\beq
\hat{C}(\epsilon) = \left(\begin{array}{cc}
\frac{K}{\sqrt{2}}\cos\epsilon & \frac{K}{\sqrt{2}}\sin\epsilon\\
-K\sin\epsilon & K\cos\epsilon
\end{array} \right)\ ,
\label{C}
\eeq
and therefore
\beq
\hat{f}_V(q^2,\epsilon) = \left(\begin{array}{cc}
\frac{m^2_\omega}{m_\omega^2-q^2}\cos^2\epsilon +
\frac{m_\phi^2}{m_\phi^2-q^2}\sin^2\epsilon &
\frac{\cos\epsilon\sin\epsilon}{\sqrt{2}}\left( \frac{m_\phi^2}
{m_\phi^2-q^2} - \frac{m^2_\omega}{m_\omega^2-q^2} \right) \\
\sqrt{2}\cos\epsilon\sin\epsilon\left( \frac{m_\phi^2}
{m_\phi^2-q^2} - \frac{m^2_\omega}{m_\omega^2-q^2}\right)  &
\frac{m_\omega^2}{m_\omega^2-q^2}\sin^2\epsilon +
\frac{m_\phi^2}{m_\phi^2-q^2}\cos^2\epsilon
\end{array} \right)\ .
\label{vd}
\eeq

Of course $\hat{f}_V(0,\epsilon)={\bf 1}$, as is required to fix
the normalization of the charges. Using eq.~(\ref{vd}) together
with eq.~(\ref{form}) we can finally write
\beq
<r^2_s>_{Dirac} = 6\left. \frac{\partial F_1^s}{\partial q^2}
\right|_{q^2=0} = 6\sqrt{2}\cos\epsilon\sin\epsilon\left
( \frac{1}{m_\phi^2} - \frac{1}{m_\omega^2} \right)F_{1i}^o(0)
+ 6\left. \frac{\partial F_{1i}^s}{\partial q^2}\right|_{q^2=0} \; ,
\label{dirac}
\eeq
\beq
<r^2_s>_{Sachs} = 6 \left. \frac{\partial }{\partial q^2}\left(
F_1^s(q^2) + \frac{ q^2}{4M_N^2}F_2^s(q^2)\right)\right|_{q^2=0} =
<r^2_s>_{Dirac} + \frac{6}{4M_N^2}F_{2i}^s(0) \; ,
\label{sachs}
\eeq
where $F_{1i}^o(0)=3/2$ \cite{ja}.

The net effect of including the physics of VMD on both, the
Dirac and Sachs square radius, is to add a contribution
$\,9\sqrt{2}\cos\epsilon\sin\epsilon\left( m_\phi^{-2} -
m_\omega^{-2} \right)\,$ to the intrinsic result. This
contribution is independent of the dynamics chosen for the
intrinsic physics. It is almost three times bigger than the
result in ref.~\cite{mu} for the Dirac strangeness radius,
and approximately eight times smaller than the result in
ref.~\cite{ja}, but with a different sign.

For the intrinsic form factor we will take a kaon loop
calculation along the lines of ref.~\cite{mu} as a reasonable
but crude model for the intrinsic physics. Our calculation
basically reproduces that of ref.~\cite{mu}; the only
difference is that
the calculations in ref.~\cite{mu} were done with $M_\Lambda$
taken to be degenerate with $M_N$ and we use the correct
$M_\Lambda - M_N$ splitting, which tends to decrease the
intrinsic strangeness radius.

The pseudoscalar meson-baryon coupling  for extended hadrons is
schematically given by
\beq
{\cal L}_{BBM} =-ig_{BBM} \bar{\Psi}\gamma_5 \Psi H(-\partial^2)
\phi \; ,
\eeq
where $\Psi$ and $\phi$ are baryon and meson fields respectively,
$H(k^2)$ is the form factor at the meson-baryon vertices and $k$
is the  momentum of the meson.

The amplitude  for the process in which the photon couples to the
baryon is
\beq
\Gamma^B_\mu(p^\prime,p) = -ig^2_{N\Lambda K} Q_\Lambda \int
\frac{d^4k}{(2\pi)^4} \Delta(k^2) H(k^2) \gamma_5 S(p^\prime,k)
\gamma_\mu S(p,k) \gamma_5 H(k^2) \; ,
\label{bv}
\eeq
where $\Delta(k^2) = (k^2-m_K^2+i\epsilon)^{-1}$ is the kaon
propagator, $S(p,k) = (\not{p}-\not{k}-M_\Lambda+i\epsilon)^{-1}$
is the $\Lambda$ propagator, $p^\prime=p+q$ with $q$ being
the photon momentum, and $Q_\Lambda$ is the $\Lambda$
strangeness charge. In a convention where the s-quark has
strangeness $+1$ we get $Q_\Lambda=1$.

The amplitude for the process in which the photon couples to
the meson is
\beq
\Gamma^M_\mu(p^\prime,p) = -ig^2_{N\Lambda K} Q_K \int
\frac{d^4k}{(2\pi)^4} \Delta((k+q)^2) (2k+q)_\mu \Delta(k^2)
H((k+q)^2) \gamma_5 S(p,k)  \gamma_5 H(k^2) \; ,
\label{mv}
\eeq
where $Q_K=-1$ is the kaon strangeness charge.

The effective baryon-meson interaction is nonlocal for extended
baryons and this induces an electromagnetic vertex current if
the photon field is present. The gauge invariance of this
effective interaction can be maintained via minimal
substitution, which generates the seagull vertex  \cite{ohta}:
\beq
i\Gamma_\mu(k,q)=\mp g_{N\Lambda K} Q_K \gamma_5(q\pm 2k)_\mu
\frac{H(k^2) - H((q\pm k)^2)}{(q\pm k)^2 - k^2} \; ,
\label{seagull}
\eeq
where the upper and lower signs correspond to an incoming or
outgoing meson respectively.

The amplitude for the process in which the photon couples to the
vertices is
\beqa
\Gamma^V_\mu(p^\prime,p)& =& -ig^2_{N\Lambda K} Q_K \int \frac{d^4k}
{(2\pi)^4} H(k^2) \Delta(k^2) \left[\frac{ (q+2k)_\mu}{ (q+k)^2-k^2}
\left(H(k^2)\, - H((k+q)^2)\right) \times \right.
\nonumber\\*[7.2pt]
& &\left.  \gamma_5 S(p-k) \gamma_5 - \frac{ (q-2k)_\mu}{
(q-k)^2-k^2} \left(H(k^2)-H((k-q)^2)\right) \gamma_5 S(p^\prime-k)
\gamma_5\right] \; .
\label{vv}
\eeqa

With these three amplitudes it is easy to check the Ward-Takahashi
identity
\beq
q^\mu(\Gamma^B_\mu(p^\prime,p) + \Gamma^M_\mu(p^\prime,p) +
\Gamma^V_\mu(p^\prime,p)) = Q( \Sigma(p) - \Sigma(p^\prime)) \; ,
\label{WT}
\eeq
where $Q$ is the nucleon strangeness $Q=Q_\Lambda + Q_K = 0$,
and $\Sigma(p)$ is the self-energy of the nucleon related to
the $K\Lambda$ loop. The sum of the three amplitudes also
guaranties the charge non-renormalization (or the Ward Identity)
\beq
(\Gamma^B_\mu + \Gamma^M_\mu + \Gamma^V_\mu)_{q=0} = Q\left(
-\frac {\partial} {\partial p^\mu} \Sigma(p)\right) = 0 \; .
\label{W}
\eeq

We finally express these amplitudes in terms of the Dirac and
Pauli nucleon intrinsic form factors
\beq
\Gamma_\mu(p^\prime,p) = \gamma_\mu F_{1i}^s(q^2) +
i\frac{\sigma_{\mu\nu}q^\nu} {2M_N} F_{2i}^s(q^2) \; .
\eeq

The numerical results for $<r^2_{is}>_{Dirac}$ and $F_{2i}^s(o)=
\mu_i^s$ ( the intrinsic strangeness magnetic moment) are
shown in Fig. 1 and Fig. 2 respectively, as a function of
the form factor cut-off $\Lambda$. The value of the coupling
and masses used are $M_N=939\MeV$, $M_\Lambda=1116\MeV$,
$m_K=496\MeV$ and $g_{N\Lambda K}/\sqrt{4\pi}=-3.944$
\cite{holz}. For the form factor at the meson-nucleon vertices
we use
\beq
H(k^2) = \frac{m_K^2 - \Lambda^2}{k^2 - \Lambda^2} \; ,
\label{fa}
\eeq
as in the Bonn potential for baryon-baryon interactions
\cite{holz}. The same form was used in ref.~\cite{mu}.
The Bonn value for the cut-off $\Lambda$ in the $N\Lambda K$
vertex is in the range 1.2 --- 1.4 $\GeV$.
We also plot in Figs. 1 and 2, for comparison, the results
of ref.~\cite{mu} where $M_N$ instead of $M_\Lambda$ was
used in the baryon propagators (in eqs.~(\ref{bv}),
(\ref{mv}) and (\ref{vv})).

In table I, we compare the results for the nucleon mean-square
strangeness radius obtained in this work  with the pure loop
calculation of ref.~\cite{mu} (using $M_\Lambda$ in the
baryon intermediate state) and with ref.~\cite{ja}. The
first two results correspond to the upper and lower Bonn
fit values of the cut-off $\Lambda$ \cite{holz}. As in
ref.~\cite{mu} the signs of our results are opposite to the
pure pole predictions. From this table we conclude that the
vector meson contribution to the Dirac strangeness radius
of the nucleon, in a world where the real $\omega$ and $\phi$
mesons  are given by eq.~(\ref{states}), is far more important
than the intrinsic contribution, even for a mixing angle as
small as $\epsilon = 0.053$. This is encouraging since, as
pointed out in ref.~\cite{mu}, the intrinsic contribution to
$<r^2_{is}>_{Dirac}$ from the seagull vertex dominates the
remaining intrinsic contributions by more than an order of
magnitude. As the prescription to determine the seagull
vertex is not unique, it is reassuring that our result,
in contrast to ref.~\cite{mu}, depends on it only weakly.
we find a Dirac strangeness radius almost four times bigger
than the pure kaon-loop calculation and around six times
smaller than Jaffe's pole result.
For the Sachs mean-square strangeness radius we get almost
twice the size of the kaon-loop calculation and around 1/3
of the pole result.

We expect that the experiments planned at CEBAF \cite{cebaf}
with the objective of constraining the nucleon's mean-square
strangeness radius will soon give some empirical basis for
deciding which picture is more appropriate to describe the
nucleon's vector current strangeness matrix element.

In conclusion, we evaluated the vector current strangeness
matrix element of the nucleon at low momentum transfer, based
on the simple physical picture of a kaon cloud, coupled to
$\omega$ and $\phi$ mesons via vector meson dominance. $\omega$
and $\phi$ are mixed states of pure $|\sbar s>$ and $(|\ubar u>
+ |\dbar d>)$. The numbers obtained in this picture for the
nucleon strangeness radius lie between the results of refs.
{}~\cite{mu} (pure kaon cloud) and \cite{ja} (pure vector meson
poles), and are both, Dirac and Sachs, of observable magnitude.
In contrast to ref.~\cite{ja}, in our model a decreasing
mixing angle decreases the strange quark matrix elements of
the nucleon. Unlike in ref.~\cite{mu}, the relative importance
of the ambiguous seagull contribution to the result is
controllable (less than 30\%).

We thank M.J. Musolf for introducing us to the problem and
for useful comments.
T.D.C.  and H.F. acknowledge support from US Dept. of Energy
under grant No. DE-FG02-93ER-40762; T.D.C. also acknowledges
support of the National Science foundation under grant No.
PHY-9058487. M.N. acknowledges the warm hospitality and
congenial atmosphere provided by the Nuclear Theory Group
of the University of Maryland and support from FAPESP-Brazil
and CNPq-Brazil.
\eject

\eject
\noindent
{\Large\bf Figure Captions}\\ \\ \\
{\bf Figure 1.} The intrinsic strangeness mean square radius of the
nucleon as a function of the cut-off $\Lambda$ in the baryon-meson
form factor. The solid and dashed lines are the results of this
work and ref.~\cite{mu} respectively.
\\ \\
{\bf Figure 2.} The intrinsic strange magnetic moment of the nucleon
as a function of $\Lambda$. Solid and dashed lines are assigned
as in fig. 1.

\eject
\noindent
{\Large\bf Table Captions}\\ \\ \\
{\bf Table I.} Theoretical results for the nucleon mean square
strangeness radius. The region of values in the first two rows
is determined by varying the cut-off in the meson-baryon vertex
form factor over the range of Bonn values \cite{holz}, 1.2 ---
1.4 $\GeV$. The results in the first row were obtained using
$M_\Lambda$ (instead of $M_N$ as in ref.~\cite{mu}) in the
baryon propagator.

\eject
\centerline{\bf Table I}
\vskip 2cm
\begin{tabular}{|c|c|c|}\hline
 & $<r^2_s>_{Dirac} (\fm^2)$ & $<r^2_s>_{Sachs} (\fm^2)$ \\
\hline & & \\ kaon-loops  & (-6.68 --- -6.90)x$10^{-3}$ &
(-2.23 --- -2.76)x$10^{-2}$ \\  &  &\\
this work  & (-2.42 --- -2.45)x$10^{-2}$ & (-3.99 ---
-4.51)x$10^{-2}$ \\ & &\\
poles (ref.~\cite{ja}) & 0.16 $\pm$ 0.06 & 0.14 $\pm$ 0.07 \\ \hline
\end{tabular}

\end{document}